\documentclass[tightenlines,showpacs,prl, floatfix,preprintnumbers,amsmath,
amssymb,twocolumn]{revtex4}

\usepackage{graphicx,color}

\usepackage{bm}

\begin{document}
\vspace{1cm}

\title{Stochastic Gravitational-Wave Background due to Primordial Binary Black Hole Mergers}

\author{Vuk Mandic$^a$, Simeon Bird$^b$, and Ilias Cholis$^b$ }
\affiliation{$^a$School of Physics and Astronomy, University of Minnesota, Minneapolis, MN 55455, USA\\
$^b$Department of Physics and Astronomy, Johns Hopkins University, Baltimore, MD 21218, USA
}
\date{\today}

\begin{abstract}
Recent Advanced LIGO detections of binary black hole mergers have prompted multiple studies investigating the possibility that the heavy GW150914 binary system was of primordial origin, and hence could be evidence for dark matter in the form of black holes. We compute the stochastic background arising from the incoherent superposition of such primordial binary black hole systems in the universe and compare it to the similar background spectrum due to binary black hole systems of stellar origin. We investigate the possibility of detecting this background with future gravitational wave detectors, and discuss the possibility of using the stochastic gravitational-wave background measurement to constrain the dark matter component in the form of black holes.
\end{abstract}


\bibliographystyle{plain}
\maketitle

\pagestyle{plain}

{\bf Introduction:}
Advanced LIGO detectors \cite{aLIGO,GW150914det} recently recorded the first two gravitational wave events: GW150914 \cite{GW150914} and GW151226 \cite{GW151226}. Both events had waveforms consistent with the mergers of binary black hole (BBH) systems, providing the first evidence that such binary systems exist and that they can merge within the lifetime of the universe. While the GW151226 system was characterized with black hole masses of $14^{+8}_{-4} M_{\odot}$ and $7^{+2}_{-2} M_{\odot}$ \cite{GW151226}, which were consistent with the dynamical black hole mass estimates in x-ray binaries, GW150914 was characterized with rather heavy individual black hole masses: $36^{+5}_{-4} M_{\odot}$ and $29^{+4}_{-4}M_{\odot}$ \cite{GW150914,GW150914PE}. Such large black hole masses are still consistent with the (relatively uncertain) high mass stellar formation process, possibly in low-metallicity environments, with binaries formed in either the field or dynamical formation scenarios \cite{belczynski_1, mapelli, spera, mandel, marchant, belczynski_2, morscher, rodriguez_1, rodriguez_2,GW150914astro}.

It has also been argued that black holes with large masses could be of primordial origin and contribute to the dark matter content of the universe \cite{bird,sasaki}. Black hole masses below $20 M_{\odot}$ have been excluded as a significant contributor to dark matter via microlensing surveys \cite{Allsman,Wyrz,Tiss}. Similarly, masses above $100 M_{\odot}$ would disrupt wide binaries \cite{Monroy,Yoo,Quinn}. While the mass range $20-100 M_{\odot}$ has also been constrained by CMB observations \cite{Ricotti:2007au, Ricotti:2007jk}, these constraints are subject to significant uncertainties, as discussed in \cite{bird}. Furthermore, it has been argued \cite{bird} that under the assumption that dark matter is made up of $\sim 30 M_{\odot}$ primordial black holes, the rate of mergers of primordial black hole binary systems would be consistent with the rate observed by Advanced LIGO detectors.

In this letter we investigate whether the origin of heavy BBH systems could be determined using the stochastic gravitational wave background measurements. We extend the study presented in \cite{bird} to compute the stochastic background as an incoherent superposition of gravitational waves emitted by all primordial black hole binaries in the universe. We investigate the possibility of detecting this background in the future runs of Advanced LIGO and of future gravitational wave detectors. Measuring this background could therefore be used to constrain the fraction of dark matter that is in the form of primordial black holes.

{\bf Merger Rate per Halo:}
We follow \cite{bird} in the computation of the local merger rate of primordial BBHs. In this model, dark matter consists of black holes of mass $M$, which directly yields the number density of such black holes in a given dark matter halo. The primordial black holes in the dark matter halo interact with each other via the emission of gravitational waves and occasionally become bound and form binaries. The cross section for this process has been computed (see Eq. 1 of \cite{bird} and \cite{Quinlan,Mouri}). Combining the black hole number density with the capture cross section and integrating over a dark matter halo yields the rate of primordial BBH mergers per halo \cite{bird}:

\begin{equation}
R_{\rm halo} (z) = \left( \frac{85\pi}{6\sqrt{2}} \right)^{2/7} \frac{2\pi}{3} \frac{G^2 M_{vir}^2 D(v) \lambda^2}{R_S^3 c g^2(C)} \left[ 1 - \frac{1}{(1+C)^3}\right].
\label{Rhalo}
\end{equation}

Here, $z$ is the redshift, $G$ is the Newton constant, $c$ is the speed of light, $\lambda$ is the fraction of dark matter energy density in the form of black holes, and we make the following definitions and assumptions:
\begin{itemize}
\item We assume the Navarro-Frank-White (NFW) halo density model, which takes the form:
\begin{equation}
\rho(r) = \frac{4\rho_s}{\frac{r}{R_S}(1 + \frac{r}{R_S})^2}
\end{equation}
where $\rho_s$ and $R_S$ are the characteristic density and radius of the halo profile, and $r$ is the distance from the center of the halo.
\item The virial radius $R_{vir}$ is defined to be the radius at which the NFW profile reaches 200 times the critical density of the universe (which is redshift dependent). The virial mass $M_{vir}$ is defined to be the mass inside the virial radius.
\item The concentration parameter is defined as $C = R_{vir}/R_S$ and is extracted from the fits to numerical simulations \cite{ludlow,prada}. Our calculations below will require extrapolating the concentration models outside of the mass and redshift range considered in simulations. To avoid divergent behavior of $C$ in such parts of the parameter space, we clip the value of $C$ at 0.5 (minimum) and 1000 (maximum). We have verified that the impact of such clipping on the estimate of the stochastic gravitational-wave spectrum is negligibly small.
\item We define the following function of concentration:
\begin{equation}
g(C) = \ln(1 + C) - \frac{C}{1+C}.
\end{equation}
\item Finally, we average over black hole velocities, using the Maxwell-Boltzmann distribution for the dark matter halo cut-off at the virial velocity $v_{vir}$:
\begin{eqnarray}
P(v,v_{dm}) & = & F_0 \left[ \exp\left(-\frac{v^2}{v_{dm}^2}\right) - \exp\left( -\frac{v_{vir}^2}{v_{dm}^2} \right) \right]\\
v_{dm} & = & \sqrt{ \frac{G M_{vir} g(C_m)}{ R_{\max} g(C)} }\\
v_{dm} & = & \frac{v_{vir}}{\sqrt{2}} \sqrt{\frac{C}{C_m} \frac{g(C_m)}{g(C)}} \\
D(v) & = & \int_0^{v_{vir}} P(v,v_{dm}) \left( \frac{2v}{c} \right)^{3/7} dv.
\end{eqnarray}
Here, $F_0$ is the normalization of the Maxwell-Boltzmann distribution, $C_m = 2.1626$, and $R_{\max} = R_S C_m$ .
\end{itemize}

\begin{figure}[!t]
\includegraphics[width=3in]{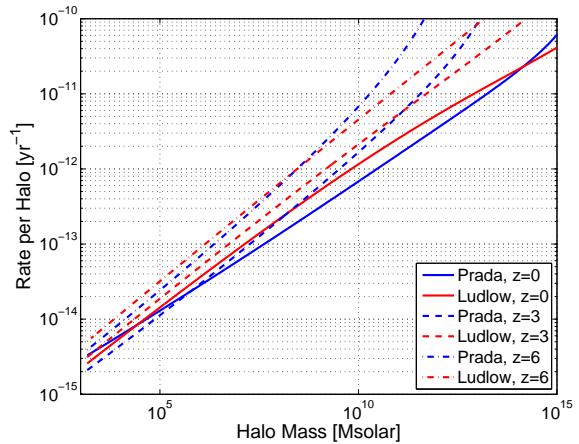}
  \caption{Primordial BBH merger rate per halo as a function of the halo virial mass for the Prada et al. \cite{prada} and Ludlow et al. \cite{ludlow} concentration models, assuming $\lambda = 1$, and for several values of redshift. The local $z=0$ curves are to be compared to the Figure 1 of \cite{bird}. }
\label{rhalo_vs_mass}
\end{figure}

Note that the merger rate per halo does not depend on the mass of the black hole (under the assumption that all black holes are of the same size). With the above definitions, we can compute the primordial BBH merger rate per halo, as a function of the halo (virial) mass. This is shown in Figure \ref{rhalo_vs_mass} which compares two concentration models, Prada et al. \cite{prada} and Ludlow et al. \cite{ludlow}, for several values of redshift. The two concentration models yield similar merger rates for the low halo masses, but the difference could be substantial at higher masses. As discussed below, the high end of the halo mass distribution contributes little to the gravitational wave spectrum, so this disagreement between concentration models (arising because of the finite box size of the underlying simulations) for high halo mass is not critical.

\begin{figure}[!t]
\includegraphics[width=3in]{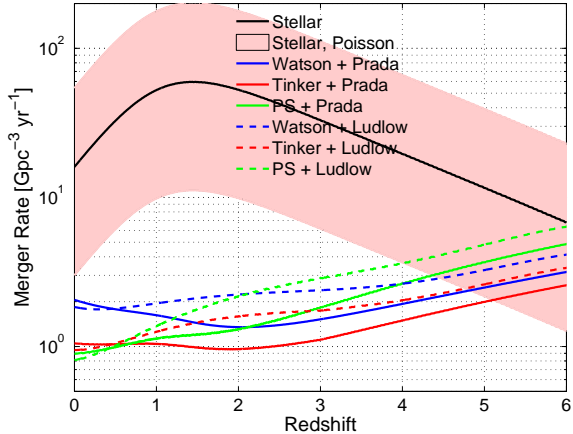}
  \caption{Primordial BBH merger rate per comoving volume as a function of redshift, using the Prada et al. \cite{prada} and Ludlow et al. \cite{ludlow} concentration models, assuming $\lambda = 1$, and for several halo mass function models \cite{watson,PS,tinker}. Note that the fiducial stellar BBH model is computed using black hole binaries which trace the cosmic star formation rate, and thus peaks around $z\sim 1-2$ \cite{GW150914stoch}. The Poisson band around the fiducial stellar model represents the statistical uncertainty in the local rate of BBH mergers \cite{GW150914stoch}. The primordial BBH merger rate in all considered models is weakly dependent on redshift and slightly increases with redshift. }
\label{rmerger_vs_z}
\end{figure}

{\bf Merger Rate per Comoving Volume:}
To calculate the merger rate per comoving volume we use the halo mass function $dn/dM_{vir}$, i.e. the number density of dark matter halos between $M_{vir}$ and $M_{vir} + dM_{vir}$, which is a function of both halo mass and redshift and has units ${\rm Gpc^{-3}}M_{\odot}^{-1}$. The halo mass function is also extracted from simulations, and we will compare several halo mass function models from the \texttt{hmf} package \cite{hmf_website}, specifically Watson et al. \cite{watson}, Press-Schechter \cite{PS}, and Tinker et al. \cite{tinker}. Several other models available within the \texttt{hmf} package were also tested to confirm that they yield results consistent with those presented here. We therefore define:
\begin{equation}
R(z) = \int R_{\rm halo}(z) \frac{dn}{dM_{vir}} dM_{vir}.
\end{equation}
Hence, $R(z)$ is the merger rate per comoving volume as a function of redshift in units of ${\rm Gpc}^{-3} \; {\rm yr}^{-1}$. Figure \ref{rmerger_vs_z} shows the merger rate obtained using several halo mass functions and two concentration models, compared to the stellar BBH model defined as the fiducial model in \cite{GW150914stoch}. Note that the primordial BBH merger rate has a very different redshift profile from the stellar one: while the stellar model follows the star formation rate (and hence peaks at redshifts 1-2), the primordial models weakly increase with redshift. Furthermore, different halo mass function models and uncertainty in the mass of the smallest halos, yield estimates of the merger rate differing by a factor of $\sim$3.

{\bf Gravitational-Wave Energy Density:} The energy density arising from an incoherent superposition of BBH systems throughout the universe has been calculated by multiple authors under the assumption that the BBH systems are of stellar origin \cite{phinney,kosenko,schneider,regfrei,regman,regreview,ros11,BBHstoch,mar11,wuCBC,GW150914stoch}. We now extend this calculation to primordial BBH systems, following the formalism of \cite{GW150914stoch}:

\begin{equation}
  \Omega_{\rm{GW}}(f)=\frac{f}{\rho_c H_0} \int_0^{z_{\rm max}}  dz \frac{R(z) \frac{dE_{\rm{GW}}(f_s,z)}{df}}{(1+z) E(\Omega_{\rm M},\Omega_{\Lambda},z)}  .
\label{eq:omega_flux}
\end{equation}
Here, $H_0$ is the present value of the Hubble parameter, $\rho_c$ is the critical energy density of the universe, and $E(\Omega_{\rm M},\Omega_{\Lambda},z) = \sqrt{\Omega_{\rm M} (1+z)^3 + \Omega_{\Lambda}}$ captures the dependence of the comoving volume on redshift for the standard flat cosmology model, with $\Omega_{\rm M} = 0.307$ and $\Omega_{\Lambda} = 1-\Omega_{\rm M}$. The energy spectrum emitted by a single binary, $dE_{\rm GW}/df$, is evaluated at the source frequency $f_s = f(1+z)$. We follow the formalism of \cite{ajith2008,BBHstoch} with the modifications from \cite{ajith2011} to calculate the contributions of the inspiral, merger, and ringdown parts of the BBH waveform. Specifically, the low-frequency part of the emitted spectrum is generated during the inspiral phase of the merger and leads to $\Omega_{\rm GW}(f) \sim f^{2/3}$, while merger and ringdown phases lead to more complex spectral behavior.

\begin{figure}[!t]
\includegraphics[width=3in]{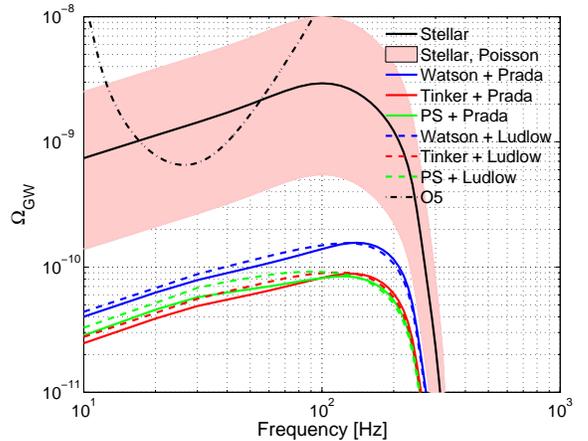}
  \caption{Gravitational-wave energy density as a function of frequency for the same models of the halo mass function and concentration as considered in Figure \ref{rmerger_vs_z} and assuming $\lambda = 1$. While different primordial models agree with each other within a factor of $\sim 2$, the fiducial stellar model is significantly louder. We note that the amplitude of the stellar fiducial model is currently uncertain due to the large errors on the local rate of BBH mergers, as denoted by the Poisson band \cite{GW150914stoch}. Also shown is the projected final sensitivity of advanced detectors, denoted O5 \cite{GW150914stoch}.}
\label{omega_vs_freq_halo}
\end{figure}

Figure \ref{omega_vs_freq_halo} shows the gravitational wave spectra for the stellar fiducial model of \cite{GW150914stoch} and for the primordial model for several combinations of halo mass functions and concentration models. All spectra assume the chirp mass of $M_{\textrm{chirp}} = 30 M_{\odot}$. For the stellar model we make the same assumptions as in \cite{GW150914stoch}: the local merger rate is taken to be $R_0 = 16 {\rm \; Gpc^{-3} \; yr^{-1}}$, and we assume that the merger rate as a function of redshift follows the star formation rate at metallicity $Z < Z_{\odot}/2$, convolved with the distribution of time-delays (between BBH formation and merger) that obeys $P(t) \sim t^{-1}$, with the minimum time delay of 50 Myr. The stellar model spectrum is characterized by a larger overall amplitude, due to both the larger local merger rate assumed in the calculation and because of the different redshift distribution of the BBH systems in the two models. It should be noted, however, that there is currently a large uncertainty in this amplitude due to the large uncertainty in the local BBH merger rate \cite{GW150914rates}. We also note that the different choices of the halo mass function or the concentration model yield variations in the primordial GW spectrum at the level of a factor of 2.

\begin{figure}[!t]
\includegraphics[width=3in]{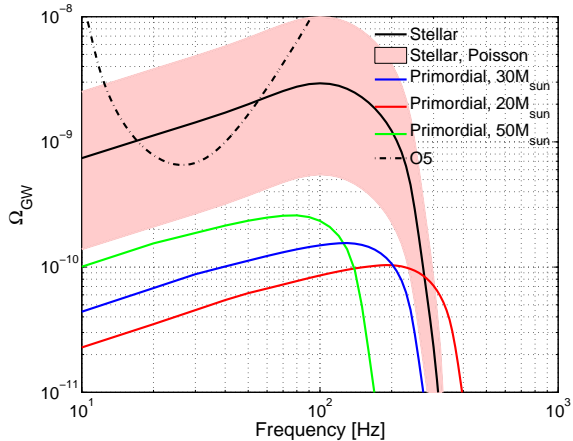}
  \caption{Gravitational-wave energy density for the primordial BBH model is shown as a function of frequency for several values of the black hole mass, assuming the Ludlow et al. concentration model \cite{ludlow} and the Watson et al. model of the halo mass function \cite{watson}. Also shown is the projected final sensitivity of advanced detectors, denoted O5, as well as the fiducial stellar model and its Poisson error band \cite{GW150914stoch}.
  }
\label{omega_vs_freq}
\end{figure}

The gravitational-wave spectrum due to BBH mergers also depends on the assumed BBH chirp mass, as discussed in \cite{GW150914stoch} for the stellar BBH systems and demonstrated for the primordial BBH systems in Figure \ref{omega_vs_freq}. Specifically, for smaller black hole masses, the binaries would merge at a higher frequency, hence shifting the entire spectrum to higher frequencies. Furthermore, since the amplitude of the spectrum scales as $M_{\textrm{chirp}}^{5/3}$, lower chirp mass results in a lower spectral amplitude at a given frequency. In reality, the BBH chirp mass distribution is likely broader than the single value assumed in our calculations above, leading to a broader peak of the spectrum.

\begin{figure}[!t]
\includegraphics[width=3in]{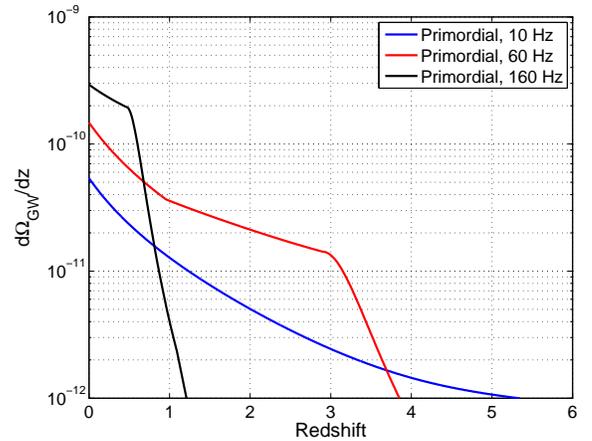}
  \caption{Gravitational-wave energy density for the primordial BBH model is shown as a function of redshift for three selected frequencies: 10 Hz, 60 Hz, and 160 Hz. For all curves we assume the Ludlow et al. concentration model \cite{ludlow} and the Watson et al. model of the halo mass function \cite{watson}. The majority of the signal is generated at redshifts below $\sim 3$, implying that the uncertainties in the concentration and halo mass function models at high redshifts have a small impact on the gravitational wave energy density spectrum.
  }
\label{omega_vs_z}
\end{figure}

The simulations used to calibrate the mass function and concentration models are most reliable at low redshift. Thus the uncertainty in the halo mass function increases for $z>4$. To examine how much this affects our results, we investigate how large a contribution high-redshift mergers make to the gravitational wave energy density spectrum. Figure \ref{omega_vs_z} shows the integrand of Eq. \ref{eq:omega_flux}, i.e. the energy density as a function of redshift for three frequencies: 10 Hz, 60 Hz, and 160 Hz. It is evident that most of the signal at the three frequencies (and hence in the sensitive frequency band of terrestrial gravitational-wave detectors) comes from redshifts smaller than $\sim 3$. Hence, lack of information about the concentration and halo mass function at high redshifts does not have a significant impact on the BBH stochastic background spectrum in the frequency band of terrestrial detectors and on our conclusions. Furthermore, as the signal is dominated by the smallest halos, which cannot form stars, baryonic effects are unlikely to change our results.

{\bf Detectability and Distinguishing Between the Stellar and Primordial Models:} Having computed the amplitude and the spectral shape of the primordial BBH stochastic background, we now investigate whether this background could be detected by future gravitational wave detectors and whether it could be distinguished from other expected stochastic backgrounds. We consider two detector pairs with one year of exposure: two Advanced LIGO detectors at their final sensitivity \cite{aLIGO,GW150914det} and two co-located detectors at the Einstein Telescope ET-D sensitivity \cite{ET,ET2}. We use the detection statistic defined by \cite{allenromano}, following past searches for the stochastic gravitational-wave background \cite{S5stoch,S6stoch}, for which the signal to noise ratio is defined as:
\begin{equation}
  \text{SNR} =\frac{3 H_0^2}{10 \pi^2} \sqrt{2T} \left[
\int_0^\infty df\>
\frac{\gamma^2(f)\Omega_{\rm GW}^2(f)}{f^6 P_1(f)P_1(f)} \right]^{1/2}\,,
\label{snr}
\end{equation}
where $H_0$ is the present value of the Hubble parameter, $T$ is the observation time (set to 1 year in our case), $\gamma(f)$ is the overlap reduction function for the chosen detector pair, arising from the different locations and orientations of the detectors \cite{allenromano}, and $P_1(f)$ and $P_2(f)$ are the strain power spectral densities of the two detectors.

Inserting the primordial BBH spectrum for $\Omega_{\rm GW}(f)$, assuming the Watson et al. halo mass function \cite{watson}, the Ludlow concentration model \cite{ludlow}, and $\lambda = 1$ (i.e. that all of dark matter is in the form of black holes) yields ${\rm SNR}<2$ for the Advanced LIGO final detector sensitivity. Hence, the primordial BBH stochastic background cannot be measured by Advanced LIGO detectors. However, a pair of co-located Einstein Telescope detectors would have sufficient sensitivity to measure the primordial BBH background with ${\rm SNR} >2$ even if the fraction of dark matter in the form of black holes is as low as $ \lambda \approx 0.2$.

Although the primordial BBH background should be strong enough to be measurable by the Einstein Telescope, distinguishing it from other types of stochastic background will be challenging. As shown in Figures \ref{omega_vs_freq_halo} and \ref{omega_vs_freq}, both primordial and stellar BBH spectra have the same frequency dependence $f^{2/3}$ in the sensitive frequency band of the Einstein Telescope. While there is currently a large uncertainty in the stellar BBH merger rate, and hence in the amplitude of the corresponding stochastic background, even the lower end of the allowed range for the merger rate yields the stochastic background spectrum that is louder than the primordial one (see \cite{GW150914stoch}). Hence, in order to detect the primordial component, the stellar component will have to be subtracted from the overall measurement, which would require a very good estimate of the local merger rate of stellar BBH systems as well as a very good understanding of the formation and evolution of these systems with time. This is challenging: as discussed in \cite{GW150914stoch}, the BBH formation mechanism (field vs dynamical), the dependence of black hole mass on metallicity at the time of black hole formation, and the distribution of the time delay between the formation and the merger of a BBH system, each lead to a factor of $\sim 2$ uncertainty in the stellar background spectral amplitude in the frequency band of terrestrial gravitational-wave detectors. In addition, both the stellar and the primordial BBH populations may be characterized by broad black hole mass distributions, implying that the corresponding spectra may be further modified relative to the spectra shown in Figures \ref{omega_vs_freq_halo} and \ref{omega_vs_freq}.
%

{\bf Conclusions:} In this paper we have computed the stochastic gravitational wave spectrum arising from the incoherent superposition of many primordial BBH mergers. We have shown that the amplitude of this spectrum is significantly lower than that arising from the stellar BBH mergers, although there is currently a large uncertainty in the local merger rate for stellar BBH systems. Our calculation is not very sensitive to the uncertainties in the primordial BBH model, such as the halo mass function and the concentration model. Consequently, the stochastic GW background measurement with Advanced LIGO detectors is unlikely to detect this background. A similar measurement with future detectors, for example Einstein Telescope, may be sufficiently sensitive to detect the primordial BBH gravitational-wave background. However, distinguishing this background from other sources of stochastic background will be challenging. Regardless, the stochastic gravitational-wave background measurements with future gravitational wave detectors may potentially yield new information on the dark matter problem.

{\it Acknowledgments:} We would like to thank Marc Kamionkowski and Eric Thrane for valuable discussions. The work of V.M. was supported in part by the NSF grant PHY-1204944 at the University of Minnesota.
S.B. was supported by NASA through Einstein Postdoctoral Fellowship Award Number PF5-160133.
I.C. was supported by NASA NNX15AB18G Grand and the Simons Foundation. LIGO-P1600241.

\end{document}